\documentclass[aps,footinbib,twocolumn,showpacs,amsmath,amssymb,superscriptaddress,prb]{revtex4}
\usepackage{graphicx,amsmath}
\usepackage{epstopdf}
\usepackage{amssymb}
\usepackage{grffile}
\usepackage[usenames]{color}
\usepackage{indentfirst}
\usepackage{float}
\usepackage{color}
\usepackage{mathrsfs}
\usepackage{dcolumn}
\usepackage{comment}
\usepackage{bm}
\usepackage[colorlinks=true,bookmarks=false,citecolor=blue,linkcolor=red,urlcolor=blue]{hyperref}

\begin{document}
\title{Simulating Composite Fermion Excitons by Density Functional Theory and Monte Carlo on a Disk}
\author{Yi Yang}
\affiliation{Department of Physics and Chongqing Key Laboratory for Strongly Coupled Physics, Chongqing University, Chongqing 401331, People's Republic of China}
\author{Songyang Pu}
\affiliation{Department of Physics and Astronomy, The University of Tennessee, Knoxville, Tennessee 37996, USA}
\author{Yayun Hu}
\email{hyy@zhejianglab.com}
\affiliation{Zhejiang Lab, Hangzhou 311100, People's Republic of China}
\author{Zi-Xiang Hu}
\email{zxhu@cqu.edu.cn}
\affiliation{Department of Physics and Chongqing Key Laboratory for Strongly Coupled Physics, Chongqing University, Chongqing 401331, People's Republic of China}

\begin{abstract}
The Kohn-Sham density functional method for the fractional quantum Hall (FQH) effect has recently been developed by mapping the strongly interacting electrons into an auxiliary system of weakly interacting composite fermions (CFs) that experience a density-dependent effective magnetic field. This approach has been successfully applied to explore the edge rescontruction, fractional charge and fractional braiding statistics of quasiparticle excitations. In this work, we investigate composite fermion excitons in the bulk of the disk geometry. By varying the separation of the quasiparticle-quasihole pairs and calculating their energy, we compare the dispersion of the magnetoroton mode with results from other numerical methods, such as exact diagonalization (ED) and Monte Carlo (MC) simulation. Furthermore, through an evaluation of the spectral function, we identify chiral ``graviton'' excitations: a spin $-2$ mode for the particle-like Laughlin state and a spin $2$ mode for the hole-like Laughlin state. This method can be extended to construct neutral collective excitations for other fractional quantum Hall states in disk geometry. 
\end{abstract}
\date{\today} 
\maketitle
\section{Introduction}
Since its discovery, the FQH effect becomes a pioneered field for exploring topological phases in strongly correlated systems~\cite{DCTHLSACG,RBL}.  Unlike the integer quantum Hall (IQH) state, the topological order of FQH states origins from electronic interactions, which gives rise to unique features such as fractional charge excitations, fractional statistics, topological ground-state degeneracy, gapless chiral edge excitations, and topological entanglement entropy, etc~\cite{topo1,topo2,topo3,topo4}. One of the most intriguing features of FQH systems is their neutral collective excitations, which have garnered significant attention due to the complex correlation physics they exhibit~\cite{YHZZH,DTS1}. The low-lying excitation spectrum in FQH systems is characterized by a magnetoroton mode, which displays a well-defined roton minimum~\cite{mini1,DTS1,mini2}. 

The pioneering work of Girvin, MacDonald, and Platzman introduced the single-mode approximation (SMA) to describe these lowest-energy neutral excitations in terms of a density wave, known as the magnetoroton. This is analogous to Feynman's theory for a superfluid helium liquid~\cite{GMP2,GMP1}. Explicit wave functions for magnetoroton have been proposed utilizing multicomponent CFs approaches and Jack polynomials~\cite{RodriguezPRB12,YHZZH}. As the wave vector $\bold{k} \rightarrow 0$, the magnetoroton mode merges into the continuum, making it difficult to observe and raising questions about its existence. Kohn's theorem significantly implies that the dipole spectral weight is entirely accounted for by the cyclotron mode, rendering the SMA entirely ineffective at zero wave vector. Consequently, long-wavelength collective excitations within a Landau level are imperceptible to electromagnetic probes in the linear response regime.

Haldane~\cite{haldane} pointed out that FQH states possess geometric degrees of freedom that are fundamental to their low-energy properties. For two-body interactions, these geometric degrees of freedom can be described by a metric that characterizes the ``area preserving'' quantum fluctuations of topological composite particles within a single Landau level. The long-wavelength colletive excitations correspond to oscillations of this geometry, possessing a spin angular momentum of 2~\cite{Jainbook, sphere2, sphere1, Ajit1,yangbograviton,PRBLY}, similar to gravitons in quantum theories of gravity~\cite{FP1,FP2,FP3}. Recent studies have demonstrated that the quench dynamics of the geometric metric can couple the ground state and the magnetoroton mode~\cite{LZ}. Moreover,  Liou \textit{et al.}~\cite{spectralfunction1} showed that these ``gravitons'' carry a definite chirality, characterized by angular momentum of either -2 or 2, depending on whether the FQH state is electron-like or hole-like. Additionally, Voinea \textit{et al.}~\cite{Fuzzysphere1,Fuzzysphere2} discovered the magnetoroton mode of $\nu=1/3$ bilayer system in transverse field is intertwined with the three-dimensional Ising conformal field theory at conformal critical points on the fuzzy sphere. During the past decade, experimentalists have made significant progress in detecting these collective modes~\cite{West1,ILS1,ILS2,ILS3,ILS4,ILS5,ILS6,phonon1,phonon2,phonon3,ILS7,ILS8,ILS9}. In particular, polarized Raman scattering experiments have revealed graviton-like excitations and their chirality through resonant inelastic scattering peaks~\cite{dulingjie1}. In general, the topological orders of FQH liquids are probed at the edge, based on the bulk-edge correspondence. However, a range of partially understood complexities at the edge complicates the interpretation of edge experiments, rendering them difficult or sometimes unclear. It was proposed ~\cite{spectralfunction3,PhysRevResearchSon} that the precise topological order of FQH liquids can be determined through their graviton mode excitations in the bulk, which deserves further investigation.

Most numerical studies of the FQH effect are performed on closed manifolds, such as the sphere and torus, using ED, density matrix renormalization group (DMRG) or trial wave functions. A major limitation of these approaches is their inability to incorporate effects such as nonuniform background confinement fields and edge physics in the simulation of a realistic two-dimensional electron gas. Disk geometry, which naturally includes a physical boundary, offers a complementary perspective, although its utility is constrained by its lowest symmetry and smaller accessible system sizes. To address these issues, a new approach based on density functional theory (DFT) has recently been developed~\cite{DFT1, DFT2,DFT3,DFTtostrong1,DFT4}. In applying DFT to FQHE, Hu \textit{et al.}~\cite{DFT1} reformulated the Kohn-Sham (KS) equations in terms of the CFs-quasiparticles that emerge as bound states of an electron and an even number of quantized vortices based on the original CF theory~\cite{Jain, Jainbook}. This CF-DFT approach accurately captures quantitative properties such as the energies, densities, fractional charges, and fractional statistics of the quasiparticles and quasiholes in the FQHE. Although CF-DFT has been effective in studying these properties, its application to neutral excitations, such as the magnetoroton mode, has not yet been explored. Representing the magnetoroton mode as a CF exciton-a pair of CF quasiparticle and quasihole-introduces significant challenges in disk geometry, particularly due to center-of-mass (COM) degeneracy~\cite{YWQ_disk}. The CF state is given by $\Psi_{\nu=n/(2pn+1)} (B)=P_{LLL}\Phi_n(B^*)\prod_{j<k}(z_j-z_k)^{2}$, where $\Phi_n(B^*)$ is the Slater determinant of CFs in an IQH state occupying $n$ effective Landau levels ($\Lambda$Ls) in an effective magnetic field $B^*$. The Jastrow factor represents vortex attachment to the CFs, and $P_{LLL}$ projects the system into the lowest Landau level. For a CF exciton, $\Phi_n(B^*)$ must include a hole in the highest occupied $\Lambda$ level and a particle in the lowest unoccupied $\Lambda$ level. Constructing CF excitons in disk geometry introduces COM degeneracy issues~\cite{YWQ_disk}. This work aims to construct CF excitons in disk geometry, thereby extending the study of excited states in this geometry. Using both DFT and MC methods, we compute the dispersion of the magnetoroton mode and compare our results with ED calculations. Additionally, by calculating the spectral function, we identify its chirality with spin-2, validating the effectiveness of our approach. Although demonstrated for the $\nu=1/3$ state, this method can be generalized to explore neutral collective excitations in a broader range of FQH states.

The remainder of this paper is organized as follow. In Sec.~\ref{sec2}, we detail the construction of CF excitons on a disk and outline our numerical methods, including DFT and MC. Sec.~\ref{sec3} focuses on analyzing the magnetoroton excitation using these techniques, while comparing the results with ED data. In Sec.~\ref{sec4}, we calculate the spectral function and confirm the presence of the chiral graviton mode excitation. Finally, we summarize our findings in Sec.~\ref{sec5}.

\section{CF Excitons in Disk Geometry}
\label{sec2}
\subsection{CF Excitons in Disk Geometry}
\label{cfexcitonindisk}
We consider a disk geometry and assume a system with rotational invariance. The single-particle orbital in a uniform magnetic field $\vec{B}=-B\hat{z}$ with the symmetric gauge is expressed as
\begin{equation}
\label{LandauWave}
	\eta_{n,m}(z,B)=\frac{(-1)^n}{l}\sqrt{\frac{n!}{2\pi 2^m (m+n)!}}\frac{z^m}{l^{m}}L_n^m(\frac{|z|^2}{2l^{2}})e^{-\frac{|z|^2}{4l^{2}}}
\end{equation}
where the particle position is given by $z=x+iy=r\exp({-im\theta})$, the effective magnetic length $l=\sqrt{\hbar c/eB}$ and $L_n^m$ is the associated Laguerre polynomial. The label $n=0,1,\cdots$ denotes the Landau level for electrons or the $\Lambda$ level for CFs, while $m=-n,-n+1,\cdots$ indicates the angular momentum.

\begin{figure}[ht]        
\center{\includegraphics[width=8cm]  {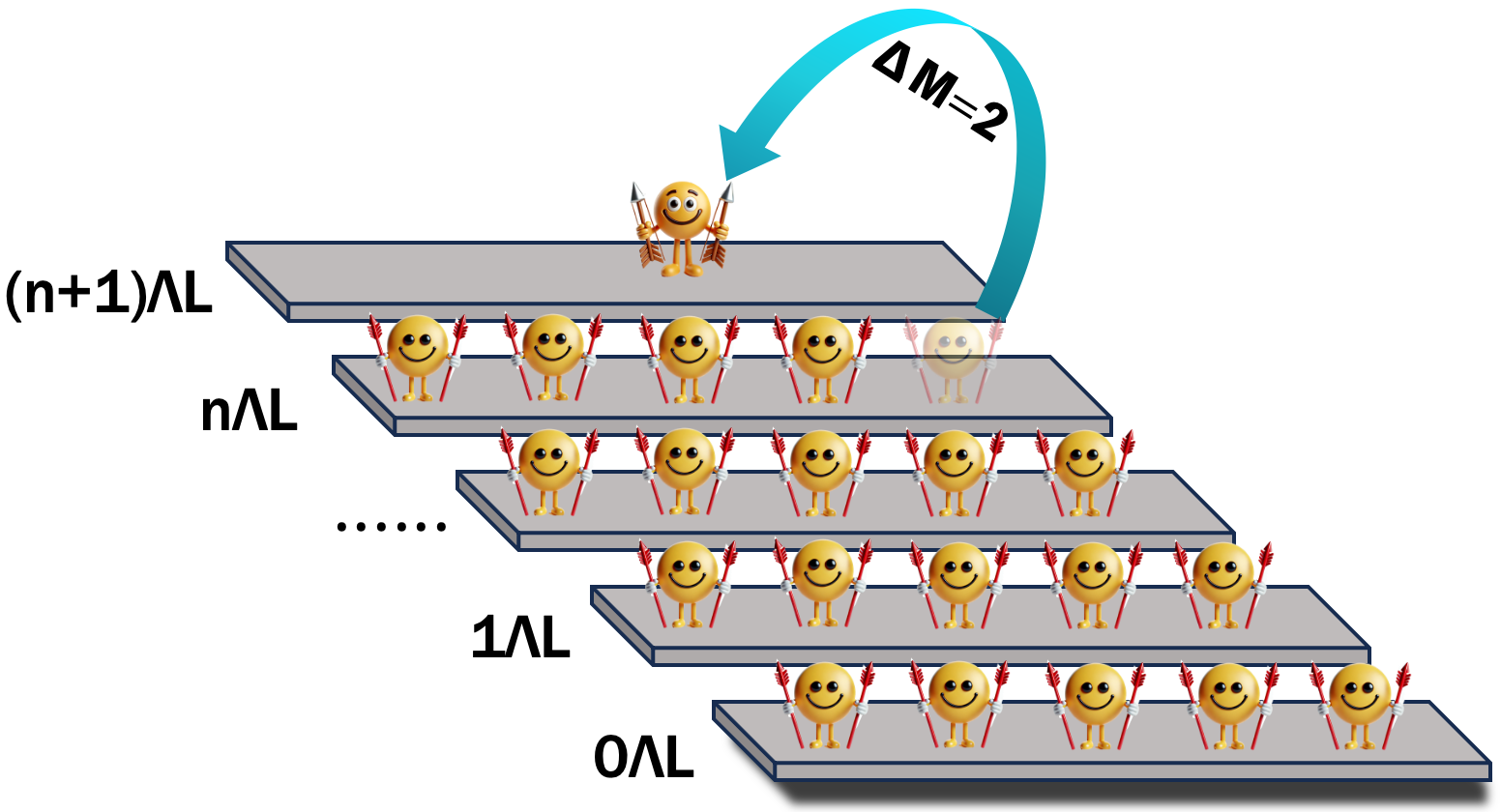}}        
\caption{\label{exciton_model} A schematic cartoon of a single CF exciton on a disk, where the stairs represent the $\Lambda$ levels of CFs. This figure depicts the excitation of a CF from the orbital with the largest angular momentum at the $n$th $\Lambda$ level to the orbital with angular momentum difference $\Delta M=2$ at the $(n+1)$th $\Lambda$ level.} 
\end{figure}

For the Jain's sequence at filling $\nu=n/(2n \pm 1)$, each electron binds to two flux quanta to form a CF, which experiences a reduced effective magnetic field $B^*=(1-2\nu)B$. The corresponding effective magnetic length is $l^*=\sqrt{\hbar c/eB^*}$. The low energy physics, including the ground states, charge excitations, and neutral excitations, can be described by CFs occupying $\Lambda$ levels in this effective magnetic field~\cite{Jainbook}. In the CF framework, a quasihole is formed by vacating an orbital in occupied $\Lambda$Ls, while a quasiparticle is constructed by occupying an orbital in the unoccupied $\Lambda$Ls. A single CF exciton, consisting of a quasiparticle-quasihole pair, is schematically illustrated in Fig.~\ref{exciton_model}. For the $\nu=1/3$ FQH state, the CFs occupy only the $0\Lambda$L. Thus, a single CF exciton is created by exciting a CF from the orbital with the maximum angular momentum in the $0\Lambda$L to an orbital with angular momentum $m$ in the $1\Lambda$L. Meanwhile, the remaining CFs occupy orbitals with angular momentum $0,1,\cdots,N_e-2$ in the $0\Lambda$L. The total angular momentum of the system is given by $M_m = \frac{(N_e-1)(N_e-2)}{2} + m + N_e(N_e-1)$, where the first two terms account for the total angular momentum of the CFs, and the third term represents the contribution of the angular momentum of the flux. Throughout this work, we use the label $\Delta M = M_{\text{gs}} - M_m$, where $M_{\text{gs}} = \frac{3N_e(N_e-1)}{2}$ is the angular momentum of the ground state. Notably, we do not impose the constraint that the exciton state is an eigenstate with zero COM angular momentum.

\subsection{DFT method for CF Excitons}
\label{dftresult}
We review the CF DFT framework and an analysis of the differences in treating CF exciton states versus ground states using DFT. According to Ref.~\cite{DFT1}, the KS equations for composite fermions are expressed as
\begin{equation}
\label{ksequation1}
	\left[T^*+V^*_{\rm KS}(\vec{r}) \right] \psi_{\alpha}(\vec{r}) = \epsilon_{\alpha} \psi_{\alpha}(\vec{r})
\end{equation}
where the KS potential is given by $V^*_{\rm KS}(\vec{r})=V_{\rm H}(\vec{r})+V_{\rm ext}(\vec{r})+V_{\rm xc}^{*}(\vec{r})+V^*_{\rm T}(\vec{r})$. The detailed forms of these potentials are provided in Appendix~\ref{DFT}. The rotational symmetry gives a conserved angular momentum. Consequently, the KS orbitals can be formally written as
\begin{equation}
	\psi_\alpha(\vec{r})=\frac{R_\alpha(r)}{\sqrt{2\pi r}}e^{-im_\alpha \theta}
\end{equation}
which could be obtained by solving Eq.~(\ref{ksequation1}) using the finite difference method. The explicit forms of $T^*$ and $V_T^*$ are derived in Appendix~\ref{KS}. For orbitals with $m_\alpha=0$, the finite difference method is less effective due to the singularity at $\vec{r}=0$. In such case, we adopt a basis expansion method, which is discussed in Appendix~\ref{m0space}. 

The CF density is computed as $ \rho(\vec{r}) = \sum_{\alpha} c_\alpha |\psi_\alpha(\vec{r})|^2$. At zero temperature, all occupied orbitals have $c_\alpha=1$, and the total number of electrons is $\sum_\alpha c_\alpha=N_e$. By combining the methods described in the Appendices~\ref{KS} and~\ref{m0space}, the KS equations in Eq.~(\ref{ksequation1}) are solved self-consistently using the standard KS-DFT iterative procedure, detailed in Appendix~\ref{iterative}. The total energy of the system consists of the following components:
\begin{equation}
   E=E_{\text{xc}}^*+E_{\text{T}}^*+E_{\text{H}}+E_{\text{ext}}+E_{\text{bb}}
\end{equation}
with
\begin{equation}
	\begin{split}
		&E_{\text{xc}}^*=\int d\vec{r} \epsilon_{\text{xc}} \rho(\vec{r}) =\int d\vec{r} \large[a(2\pi)^{1/2}\rho^{3/2}+(2b-f)\pi\rho^2+g\rho \large],\\
		&E_{\text{T}}^*= \sum_\alpha \langle \psi_\alpha|T^*|\psi_\alpha\rangle,\\
		&E_{\text{H}} =  \frac{1}{2} \int d\vec{r} \int d\vec{r}'\frac{\rho(\vec{r}')\rho(\vec{r})}{|\vec{r}-\vec{r}'|},\\
		&E_{\text{ext}}= \int d\vec{r} V_{\rm ext} \rho(\vec{r}),\\
		&E_{\text{bb}}= \frac{8N_e}{3\pi}\sqrt{\frac{N_e}{6}}.
	\end{split}
\end{equation}
here, $E_{\text{H}},~E_{\text{ext}},~\rm{and}~E_{\text{bb}}$ represent the electron-electron, electron-background, and background-background interactions, respectively. $E_{\text{T}}^*$ is the kinetic energy of the CFs, and $E_{\text{xc}}^*$ is the exchange correlation energy, which corrects the discrepancies between the auxiliary system and the real system in terms of kinetic energy and interaction energy. For a homogeneous ground state, only $E_{\text{xc}}^*$ contributes in the thermodynamic limit, as all other terms cancel out, i.e., $\lim_{N_e\rightarrow \infty}E/N_e=E_{\mathrm{xc}}/N_e$.

For the excited state at $\nu=1/3$, we determine the energy of CF excitons by traversing the quastiparticle positions in Fig.~\ref{exciton_model} and solving the corresponding KS equations. Specifically, the creation of a CF exciton involves removing a CF from the lowest $\Lambda$ level with angular momentum $N_e-1$ and placing it into the first $\Lambda$ level with angular momentum $m$. The remaining CFs occupy the lowest $\Lambda$ level with angular momentum ranging from 0 to $N_e-2$.

In addition to the conventional DFT scheme described above, such treatment of excitations can also be justified through the so-called constrained DFT formalism~\cite{constrainedDFT}, whose accuracy would be determined in principle by the accompanying exchange correction energy in use.

\subsection{MC method for CF Excitons}
\label{mcresult}
The ground state wave function of $\nu=1/3$ is constructed as described in Refs.~\cite{Jain,wavefunction1,Jainbook,wavefunction3,wavefunction4,wavefunction5}
\begin{widetext}
\begin{equation}
\label{gsmc}
\Psi_{\text{1/3}}=
\begin{vmatrix}
\eta_{0,0}(z_1,B^*) & \eta_{0,0}(z_2,B^*) & \eta_{0,0}(z_3,B^*)& \cdots \\
\eta_{0,1}(z_1,B^*) & \eta_{0,1}(z_2,B^*) & \eta_{0,1}(z_3,B^*)& \cdots \\
\eta_{0,2}(z_1,B^*) & \eta_{0,2}(z_2,B^*) & \eta_{0,2}(z_3,B^*)& \cdots \\
\cdots & \cdots & \cdots & \cdots \\
\eta_{0,N_e-1}(z_1,B^*) & \eta_{0,N_e-1}(z_2,B^*) & \eta_{0,N_e-1}(z_3,B^*) & \cdots \\
\end{vmatrix}
[\prod_{j<k} (z_j-z_k) e^{-\sum_{j}\frac{|z_j|^2}{4l_1^2}}]^{2}
\end{equation}
\end{widetext}
where $l^*=\sqrt{3}l$, $l_1$ is the magnetic length at $\nu=1$, and, according to CF theory, we have $1/l^{*2} + 2/l_1^{2} = 1/l^2$. At $\nu=1/3$, it follows that $l_1 = l^*$. Inspired by the quasiparticle and quasihole wave functions, we construct the trial wave function for a single CF exciton as:
\begin{widetext}
\begin{equation}
\Psi_{m}^{\text{1/3,exciton}}=P_{LLL}
\begin{vmatrix}
\eta_{1,m}(z_1,B^*) & \eta_{1,m}(z_2,B^*) & \eta_{1,m}(z_3,B^*)& \cdots \\
\eta_{0,0}(z_1,B^*) & \eta_{0,0}(z_2,B^*) & \eta_{0,0}(z_3,B^*)& \cdots \\
\eta_{0,1}(z_1,B^*) & \eta_{0,1}(z_2,B^*) & \eta_{0,1}(z_3,B^*)& \cdots \\
\cdots & \cdots & \cdots & \cdots \\
\eta_{0,N_e-2}(z_1,B^*) & \eta_{0,N_e-2}(z_2,B^*) & \eta_{0,N_e-2}(z_3,B^*) & \cdots \\
\end{vmatrix}
[\prod_{j<k} (z_j-z_k) e^{-\sum_{j}\frac{|z_j|^2}{4l_1^2}}]^{2}
\end{equation}
\end{widetext}
Here $m$ denotes the angular momentum of the quasiparticle in the $1\Lambda$L. We employ the standard Jain-Kamilla (JK) projection method~\cite{LLL1,LLL2}, a computationally efficient approximate method for implementing the LLL projection.

Upon projection, the trial wave function for the exciton becomes: 
\begin{widetext}
\begin{equation}
\label{excitonmc}
\Psi_{m}^{\text{1/3,exciton}}=
\begin{vmatrix}
P_{LLL}[\eta_{1,m}(z_1,B^*)J_1] & P_{LLL}[\eta_{1,m}(z_2,B^*)J_2] & P_{LLL}[\eta_{1,m}(z_3,B^*)J_3]& \cdots \\
\eta_{0,0}(z_1,B^*)J_1 & \eta_{0,0}(z_2,B^*)J_2 & \eta_{0,0}(z_3,B^*)J_3& \cdots \\
\eta_{0,1}(z_1,B^*)J_1 & \eta_{0,1}(z_2,B^*)J_2 & \eta_{0,1}(z_3,B^*)J_3& \cdots \\
\cdots & \cdots & \cdots & \cdots \\
\eta_{0,N_e-2}(z_1,B^*)J_1 & \eta_{0,N_e-2}(z_2,B^*)J_2 & \eta_{0,N_e-2}(z_3,B^*)J_3& \cdots \\
\end{vmatrix} e^{-\sum_{j}\frac{|z_j|^2}{2l_1^2}}
\end{equation}
where $J_N = \prod_{j\neq N}(z_N-z_j)$, and
\begin{equation}
		P_{LLL}[\eta_{1,m}(z_N,B^*)J_N]=\frac{(l^{*2}-1)(m+1)z_N^m-z_N^{m+1}\sum_{j\neq N}\frac{1}{z_N-z_j} }{l^{*m+3}}\prod_{k\neq N}(z_N-z_k) \sqrt{\frac{1}{2\pi 2^m (m+1)!}} e^{-\frac{|z_N|^2}{4l^{*2}}} 
\end{equation}
\end{widetext}

We utilized the Metropolis MC algorithm~\cite{Metropolis,backpotential,yangyi} to sample these wave functions.  The total energy of the system is composed of three components:
\begin{equation}
	\begin{split}
		E &= E_{\text{ee}}+E_{\text{eb}}+E_{\text{bb}}, \\
		E_{\text{ee}} &= \frac{\langle \Psi_m^{\text{1/3,exciton}}|\sum_{i<j} \frac{1}{|z_i-z_j|}| \Psi_m^{\text{1/3,exciton}} \rangle}{\langle \Psi_m^{\text{1/3,exciton}}| \Psi_m^{\text{1/3,exciton}} \rangle}, \\
		E_{\text{eb}} & = \frac{\langle \Psi_m^{\text{1/3,exciton}}|\sum_{i=1}^{N_e}V_{\rm ext}(z_i)| \Psi_m^{\text{1/3,exciton}} \rangle}{\langle \Psi_m^{\text{1/3,exciton}}| \Psi_m^{\text{1/3,exciton}} \rangle}, \\
        E_{\text{bb}}&=\frac{8N_e}{3\pi}\sqrt{\frac{N_e}{6}}.
	\end{split}
\end{equation}

Here, the potential $V_{\rm ext}(z)$ is consistent with the definition provided in the DFT section. All results presented in this paper are obtained by discarding 1,000,000 thermalization MC samples and using approximately $10^5$ MC samples for statistical averaging.

\subsection{Ground state energy per particle}
\begin{figure}[ht]        
\center{\includegraphics[width=9cm]  {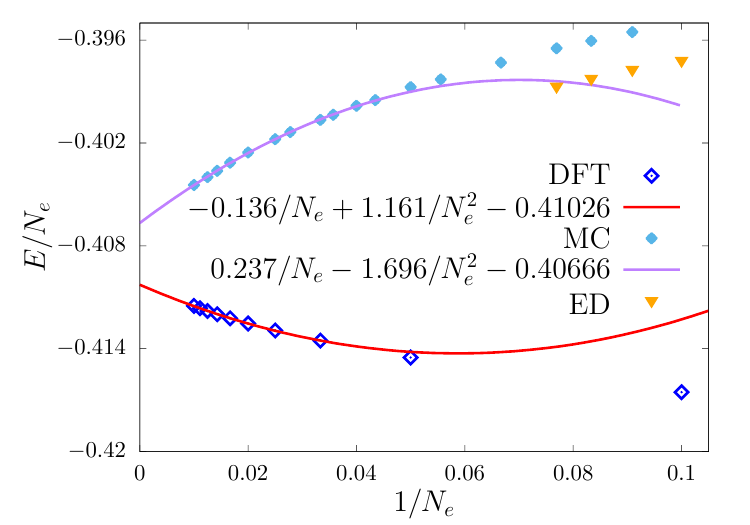}}        
\caption{\label{dft_gs_energy} The ground-state energy per electron at $\nu=1/3$. Different symbols represent the numerical results obtained from DFT, MC, and ED methods, while the lines denote the least squares fits to the data. The thermodynamic limit values obtained from DFT ($-0.41026 \pm 0.00003$) and MC ($-0.40666\pm 0.00006$) agree well with the previously reported value of $\epsilon \approx -0.41015$~\cite{backpotential,gsenergy2,gsenergy1}. The ED results are computed using the Coulomb interaction as described in Refs.~\cite{Jainbook, Coulomb1, Coulomb}.} 
\end{figure}

We calculate the ground state energy for systems of various sizes at $\nu=1/3$ and perform an extrapolation to the thermodynamic limit. These states are constructed by placing the CFs at the lowest $\Lambda$ level, with angular momentum ranging from $0$ to $N_e-1$. The computed results are shown in Fig.~\ref{dft_gs_energy}. The extrapolated energy per particle obtained from DFT, $-0.41026 \pm 0.00003$, and MC, $-0.40666\pm 0.00006$, align well with the value $\epsilon \approx -0.41015$ reported in previous studies by other methods~\cite{backpotential,gsenergy2,gsenergy1}. It is worth mentioning that the DFT calculations incorporate Coulomb interactions using the local density approximation (LDA), which results in a lower energy than the realistic Coulomb interaction. This is evident from the fact that the DFT points in Fig.~\ref{dft_gs_energy} lie below the ED points.

Using the configurations of the excited states and trial wave functions, we are then able to calculate the density and energy of the magnetoroton mode. This approach allows us to explore the intricate properties of FQH states, particularly focusing on the collective excitations that play a crucial role in understanding the system's dynamics and topology.

\section{Magnetoroton Dispersion}
\label{sec3}
\begin{figure}[ht]        
\center{\includegraphics[width=9cm]  {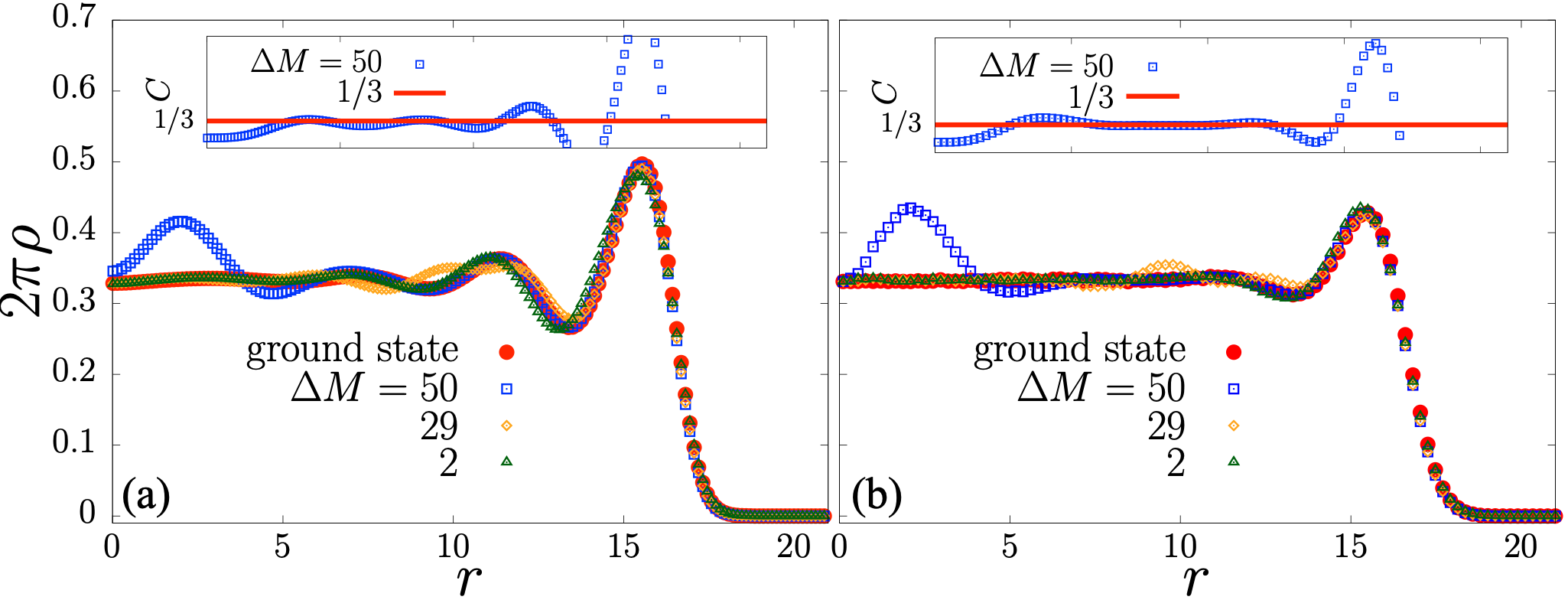}}        
\caption{\label{density} CF exciton density profiles calculated using DFT (a) and MC (b) for $N_e = 50$. Three distinct exciton densities are shown in both panels. The insets illustrate the charge accumulation $\mathcal{C}=\int dr (1/3-2\pi\rho)r$ for the largest $\Delta M$, corresponding to the creation of a quasiparticle with charge $e/3$ at the center of the disk. Notable differences are observed, particularly at the edges. The DFT results exhibit stronger density oscillations due to Coulomb interactions, while the MC method generates smoother density profiles, consistent with the trial wavefunctions.} 
\end{figure}

Based on the methods outlined in Sections \ref{cfexcitonindisk}, \ref{dftresult}, and \ref{mcresult}, we construct CF exciton states and analyze their density profiles and energy dispersion. We first present the density profiles of several exciton states. The results are shown in Fig.~\ref{density}. The inset illustrates the charge accumulation, defined as $\mathcal{C} = \int dr (1/3 - 2\pi\rho)r$, for the largest $\Delta M$. This indicates the formation of a quasiparticle carrying a fractional charge $e/3$ at the center of the disk. Notably, differences in density profiles are observed between the two numerical methods, particularly at the edges. The DFT results, which incorporate Coulomb interactions, exhibit more pronounced density oscillations compared to the relatively smoother density profile produced by the MC method, which corresponds to model wavefunctions.

\begin{figure}[ht]        
\center{\includegraphics[width=9cm]  {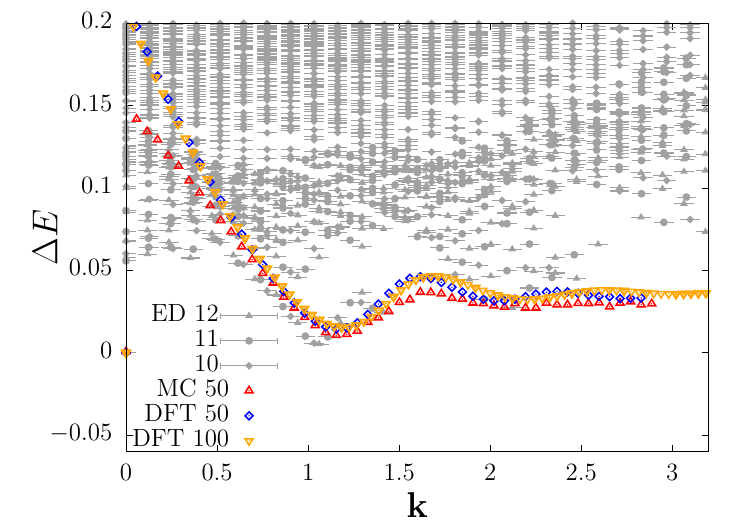}}        
\caption{\label{exciton_energy} CF exciton energy calculated using three different methods. The first half of each legend entry indicates the numerical method, and the second half denotes the system size. All methods exhibit the same dispersion trend, with consistent energy gaps. The horizontal axis is $\bold{k}=\Delta M / \sqrt{6N_e}$, allowing data from different system sizes to collapse onto a single curve for each method. For example, the DFT results for 50 and 100 electrons are in excellent agreement. Moreover, the roton minimum positions are also consistent across all methods. The discrepancy between DFT and MC in the long-wavelength limit arises from DFT's inability to enforce the LLL projection. Finally, the energy gap with Coulomb interaction is smaller than that of the model interaction (e.g., GMP mode), which is consistent with results by Jain \textit{et al.}~\cite{jainsphereenergy} in CF diagonalization on the sphere. For example, Fig.~2 in~\cite{jainsphereenergy} demonstrates that the neutral gap for the $\nu=1/3$ state is approximately 0.008.}
\end{figure}

Next, we calculate the energy of the exciton states depicted in Fig.~\ref{exciton_energy}, along with those obtained from the ED with Coulomb interactions for comparison. The horizontal axis is defined as $\bold{k}=\Delta M / \sqrt{6N_e}$, where $\sqrt{6N_e}$ is the radius of the disk. In this case, we find that data of different sizes from the same numerical method collapse into a single dispersion curve. As an example, we show the roton energies for systems with 50 and 100 electrons using DFT method. The results from all three numerical methods—DFT, MC, and ED—exhibit consistent trends, including an identical roton minimum and energy gap. However, the limited system size in the ED calculations restricts the accurate determination of the charge gap, although the neutral gap is estimated to be approximately $0.008$. As $\Delta M$ increases, the energy spectrum flattens, corresponding to a greater spatial separation between the quasihole and quasiparticle. In this regime, the energy gap represents the charge gap.

Furthermore, the similar trend in energy dispersion between the DFT and the MC methods may be attributed to a correspondence between the matrix elements of FQHE and IQHE, as noted in Ref.~\cite{wavefunction1}. In particular, for a single $\delta$ impurity and single excitation in an otherwise clean system, a numerical similarity between FQHE and IQHE matrix elements has been observed. For multi-excitation scenarios, Ref.~\cite{wavefunction1} suggests that a self-consistent treatment of the effective magnetic field for the IQHE of CFs via DFT is necessary, as we have implemented in our work. This nontrivial correspondence implies a deeper connection between the IQHE and FQHE, and it is expected to hold in the presence of weak disorder. Despite this similarity, notable discrepancies between the DFT and MC results emerge as $\bold{k}$ decreases, primarily due to DFT's inability to enforce the LLL projection. This discrepancy is also evident in the density profiles as shown in Fig.~\ref{density}, where density variations are most pronounced at the system's edge in the long-wavelength limit.

These analyses demonstrate that DFT and MC methods provide complementary insights into the density and energy characteristics of CF excitons. Moreover, the systematic comparison with ED results underscores the robustness of these numerical techniques for studying collective excitations in FQH systems.

\section{Spectral Function Analysis}
\label{sec4}
In this section, we explore the ``graviton" excitation with spin-2 by examining the long-wavelength limit of neutral collective excitations in FQH systems. According to Haldane's geometric description, these long-wavelength collective excitations in the FQH state appear as oscillations of its intrinsic geometric metric. Notably, the quantum of these oscillations carries a spin angular momentum of 2, similar to the gravitons in quantum gravity theory. Moreover, it was found that these ``gravitons” carry a definitive chirality (or angular momentum) which is either $-2$ or $+2$, depending on whether the FQH liquid is electron-like or hole-like. This was recently observed in the spectra of the polarized Raman scattering process~\cite{dulingjie1}. Numerically, this excitation is evident as resonance peaks in the spectral function~\cite{spectralfunction1,spectralfunction3,spectralfunction2} which is defined as
\begin{equation}
\label{spectral}
	I_\sigma(M) = \sum_m |\langle \Psi_{m}^{\text{1/3,exciton}} |\hat O_\sigma|\Psi_{1/3} \rangle |^2 \delta(M-M_m)
\end{equation}
where $|\Psi_{1/3} \rangle $ denotes the ground state at $\nu=1/3$, and $\Psi_{m}^{\text{1/3,exciton}}$ represents the CF exciton states with total angular momentum $M_m$, as defined in Eq.~(\ref{excitonmc}). This spectral function serves as an analog to the oscillating metric of gravitational waves. We evaluate the spectral function as a function of the total angular momentum for the full CF exciton branch rather than as a function of energy for all eigenstates, which makes Eq.~(\ref{spectral}) different from its original definition in Ref.~\cite{spectralfunction1}. In disk geometry, the chiral nature of the ``graviton'' excitations can be characterized by the following operators: 
\begin{equation}
	\begin{split}
		\hat O_{+} &= \sum_M |m+2,M\rangle \langle m,M|, \\
		\hat O_{-} &= \sum_M |m,M\rangle \langle m+2,M|.
	\end{split} 
\end{equation}
where $|m,M\rangle$ denotes a two-body state with COM angular momentum $M$ and relative angular momentum $m$.

In this work, we focus solely on the fermionic Laughlin state, for which $m=1$. The specific actions of these operators are as follows: $\hat O_{-}$ creates excitations with angular momentum $-2$, corresponding to the angular momentum of the ``graviton" mode. In contrast, $\hat O_{+}$ creates excitations with angular momentum $2$ by converting a pair of paricles with relative angular momentum $m$ into $m+2$. This process annihilates the $\nu=1/3$ Laughlin state, causing $I_+(M)$ to vanish. 

\begin{figure}[ht]        
\center{\includegraphics[width=8.6cm]  {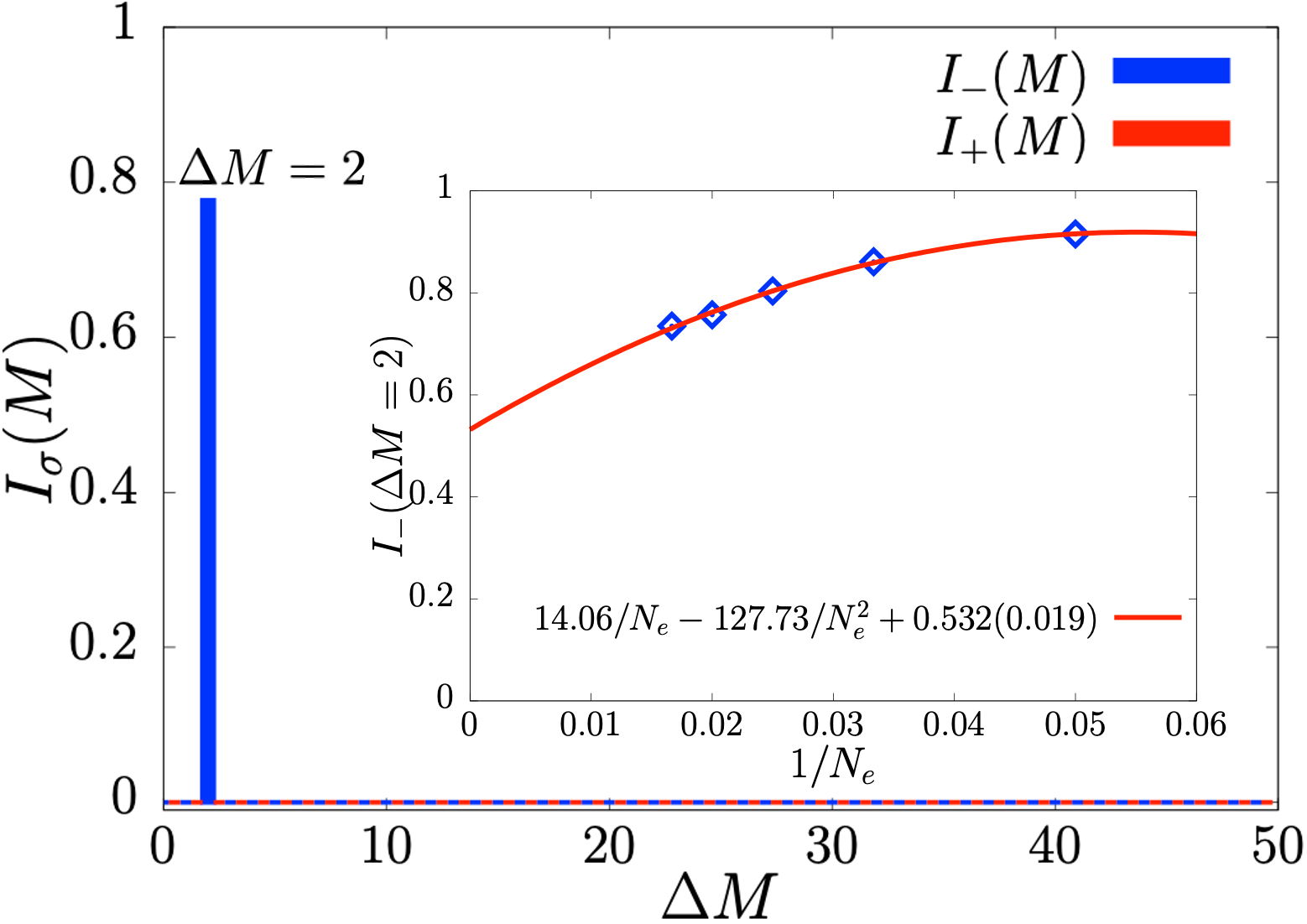}}        
\caption{\label{spectralfunction} Spectral function obtained from MC simulations for $N_e = 50$ electrons. To improve clarity, the horizontal axes of $I_+(M)$ and $I_-(M)$ are shifted by 0.1 units to the right and left, respectively. The distinct peak in $I_-(M)$ at $\Delta M = 2$ confirms the presence of a chiral graviton mode with spin $-2$. The inset shows the extrapolation of the peak values of $I_-(M)$ at $\Delta M = 2$ to the thermodynamic limit, with the red line representing the least squares fit. As the system size increases, the peak values decrease due to the greater continuity of the excitation spectrum. Consequently, the graviton mode spans a broader range of states, reducing the maximum peak amplitude.} 
\end{figure}

Using MC simulations, We calculate the chiral spectral functions $I_+(M)$ and $I_-(M)$, with $|\Psi_{1/3}|^2$ as the sampling function. We then perform a thermodynamic limit extrapolation of peak values for different system sizes. The results are normalized by the factor $\langle \Psi_{1/3}| \hat O^\dagger_{\sigma} \hat O_{\sigma}|\Psi_{1/3} \rangle$. As shown in Fig.~\ref{spectralfunction}, $I_-(M)$ exhibits a pronounced peak at $\Delta M=2$, indicating the presence of a spin $-2$ ``graviton" excitation. In contrast, all other positions of $I_-(M)$ and all values of $I_+(M)$ are zero. As the number of electrons increases, the graviton mode in the $\Delta M = 2$ space involves a greater number of excited states, leading to a reduction in the peak amplitude.

We also calculate the spectral function for the $\nu = 2/3$ state, which is the particle-hole conjugate of the $1/3$ state and exhibits opposite chirality. This $2/3$ state can be constructed by considering composite fermions at filling factor $\nu^* = n + 1$ in a negative effective magnetic field~\cite{Ajit2_3}. The corresponding trial wavefunctions for its excitons are given by

\begin{widetext}
\begin{equation}
\label{excitonmc2_3}
\Psi_{m}^{\text{2/3,exciton}}=P_{LLL}
\begin{vmatrix}
\eta_{0,0}^*(z_1,B^*)J(z_1) & \eta_{0,0}^*(z_2,B^*)J(z_2)  & \eta_{0,0}^*(z_3,B^*)J(z_3)  & \cdots \\
\cdots & \cdots & \cdots & \cdots \\
\eta_{0,N_e/2-1}^*(z_1,B^*)J(z_1) & \eta_{0,N_e/2-1}^*(z_2,B^*)J(z_2) & \eta_{0,N_e/2-1}^*(z_3,B^*)J(z_3) & \cdots \\
\eta_{1,-1}^*(z_1,B^*)J(z_1) & \eta_{1,-1}^*(z_2,B^*)J(z_2)  & \eta_{1,-1}^*(z_3,B^*)J(z_3) & \cdots \\
\cdots & \cdots & \cdots & \cdots \\
\eta_{1,N_e/2-3}^*(z_1,B^*)J(z_1) & \eta_{1,N_e/2-3}^*(z_2,B^*)J(z_2)  & \eta_{1,N_e/2-3}^*(z_3,B^*)J(z_3) & \cdots \\
\eta_{2,m}^*(z_1,B^*)J(z_1) & \eta_{2,m}^*(z_2,B^*)J(z_2)  & \eta_{2,m}^*(z_3,B^*)J(z_3) & \cdots \\
\end{vmatrix}  e^{-\sum_{j}\frac{|z_j|^2}{2l_1^2}}
\end{equation}
\end{widetext}

\begin{figure}[ht]        
\center{\includegraphics[width=9cm]  {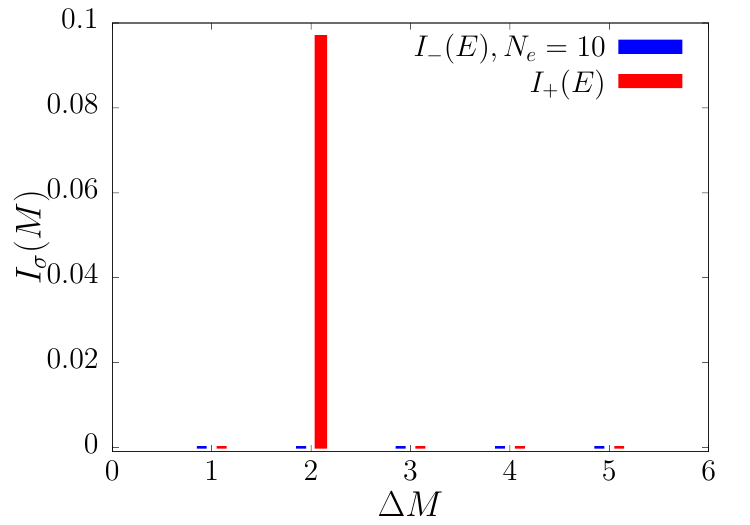}}        
\caption{\label{spectralfunction2_3} Spectral function obtained from MC simulations for $N_e=10$ electrons at $\nu=2/3$. The horizontal axes of $I_+(M)$ and $I_-(M)$ are shifted by 0.1 units to the right and left, respectively. A distinct peak in $I_+(M)$ at $\Delta M = 2$ confirms the presence of a chiral graviton mode with spin $+2$, with chirality opposite to that of the $\nu=1/3$ state.} 
\end{figure}

We apply the JK projection method, the same as for the $\nu = 1/3$ state. However, due to the computational complexity of $P_{LLL}$, which involves taking $N_e$th derivatives, we present results only for small systems. The outcomes are shown in Fig.~\ref{spectralfunction2_3}. A unique peak in $I_+(M)$ at $\Delta M = 2$ confirms the presence of a spin $+2$ graviton mode excitation with opposite chirality compared to the $1/3$ state.

\section{Summaries and Discussions}
\label{sec5}
We have developed a method for constructing CF excitons in disk geometry for FQH state at $\nu = 1/3$. By calculating the magnetoroton mode with both DFT and MC techniques and comparing them to ED, we have demonstrated consistent roton minimum and energy gaps across all three methods. The overall similarity in the trend of exciton dispersion between the DFT and MC methods is expected, given the nontrivial correspondence between the IQHE of CFs and the FQHE of electrons. However, the significant discrepancy observed in the long-wavelength limit between DFT and MC is attributed to the inability of DFT to enforce the LLL projection. Furthermore, our calculations of the spectral function reveal a clear peak in $I_-(M)$, confirming the presence of a spin $-2$ ``graviton'' excitation. We also show that the $2/3$ state hosts a graviton mode with opposite chirality compared to the $1/3$ state. 

This method is extendable to the entire Jain series and can accommodate excitations involving higher $\Lambda$ levels, such as spin-4 CF excitons. Furthermore, we note that the state we considered aligns with the scenario described in a recent manuscript~\cite{Ajit1}, where the equivalence between GMP and CF ``graviton'' descriptions is proposed. Furthermore, this approach can be generalized to the study of parton excitons by fractionalizing the electron charge, enabling the exploration of more intricate structures and providing a comprehensive framework for investigating a wide range of FQH excitations.

Finally, as highlighted in \cite{yangbograviton}, unlike model interactions, the collective excitations under Coulomb interaction will likely involve multiple chiral graviton modes. This complexity means that the spectral function may exhibit numerous resonance peaks, posing significant challenges to both theoretical analysis and numerical simulations. Addressing these challenges and refining our understanding of these resonance structures is a promising direction for future research.

\acknowledgments
 Yi Yang thanks Yuzhu Wang, Ying-Hai Wu, Tongzhou Zhao, and Xin Wan for valuable discussions during the FQHE-2024 workshop, and to Ajit C. Balram for insightful email exchanges. Special thanks to Jainendra K. Jain for providing profound insights, conveyed through Songyang Pu. This work was supported by National Natural Science Foundation grants Nos. 12474140 and 12347101, the Chongqing Research Program of Basic Research and Frontier Technology Grant No. cstc2021jcyjmsxmX0081 and the Fundamental Research Funds for the Central Universities Grant No. 2024CDJXY022. Yayun Hu was supported by Natural Science Foundation grants No. 12204432.

\appendix
\section{Detailed Methodology for CF KS-DFT Calculations} 
\label{DFT}
\begin{widetext}
\begin{equation}
\label{ksequation2}
	\begin{split}
		T^*&=\frac{1}{2m^*}\left(\vec{p}+\frac{e}{c}\vec{A}^*(\vec{r})\right)^2 \\
		V_{\rm H}(\vec{r})&=\int d\vec{r}'\frac{\rho(\vec{r}')}{|\vec{r}-\vec{r}'|}\\
		V_{\rm ext}(\vec{r}) &= \left\{ 
		\begin{array}{ll}
        -\sqrt{\frac{2}{3}N_e} \frac{2E(\frac{r^2}{6N_e})}{\pi} ,\frac{r}{\sqrt{6N_e}}\leq  1 \\     
        -\sqrt{\frac{2}{3}N_e} \frac{_2F_1(\frac{1}{2},\frac{1}{2};2;\frac{6N_e}{r^2})}{2r/\sqrt{6N_e}} ,\frac{r}{\sqrt{6N_e}}\geq 1 \\
        \end{array} \right.  \\
		V_{\rm xc}^*(\vec{r})&=\frac{\delta E^*_{\text{xc}}}{\delta \rho}=\frac{3}{2}a[2\pi l^2 \rho(\vec{r})]^{\frac{1}{2}}+(2b-f)[2\pi l^2 \rho(\vec{r})]+g\\
		V^*_{\rm T}(\vec{r})&=\sum_{\alpha}c_{\alpha}\langle\psi_{\alpha}|\frac{\delta T^*}{\delta \rho(\vec{r})}|\psi_{\alpha}\rangle
	\end{split}
\end{equation}
\end{widetext}
The parameters for the LDA exchange-correlation energy are $a=-0.78213,b=0.2774,f=0.33,g=-0.04981$. The effective vector potential is given by $\vec{A}^*(r)=\frac{1}{ r}\int_0^r r' B^*(r')dr' \vec{e}_{\theta}$, where $\nabla \times \vec{A}^*(r)=B^*(r)\vec{e}_{\rm z}=\left[ B-2\rho(r)\phi_0\right]\vec{e}_{\rm z}$. Here, $\rho_{\text{b}}=\nu_0/2\pi l^2$ denotes the uniform background charge density distributed over a disk of radius $R_{\text{b}}$, such that $\pi R_{\text{b}}^2 \rho_{\text{b}}=N_e$. For the $1/3$ Laughlin state, $\nu_0=1/3$. The distance between the background layer and the electron liquid is denoted by $d$, and in this work, we set $d=0$. The parameter $\alpha$ represents both the angular momentum and the energy level of the single-particle orbital. The external potential $V_{\rm ext}(\vec{r})$ is computed as described in~\cite{backpotential}. 

\section{Details of $T^*$ And $V_T^*$} 
\label{KS}
The normalized KS orbitals are given by
\begin{equation}\label{variablesepperation}
  \psi_{\alpha}(\vec{r})=\frac{{R}_{\alpha}(r)}{\sqrt{2\pi r}}\exp(-im_{\alpha}\theta).
\end{equation}
In polar coordinates, the Hamiltonian $\mathcal H^*=T^*+V^*_{\rm KS}(r)$ in Eq.~(\ref{ksequation1}) and Eq.~(\ref{ksequation2}) is:
\begin{widetext}
\label{ksStarpolarform}
    \begin{equation}
    	\begin{split}
    		\mathcal H^*&=\frac{1}{2m^*}\left(\vec{p}+\frac{e}{c}\vec{A}^*(\vec{r})\right)^2+V^*_{\rm KS} =\frac{1}{2m^*}\bigg[-\hbar^2\big(\frac{\partial^2}{\partial r^2}+\frac{1}{r^2}\frac{\partial^2}{\partial \theta^2}+\frac{1}{r}\frac{\partial}{\partial r}\big)+\frac{\hbar e}{c}\mathcal{B}(r)\frac{\hat{L}_z}{\hbar}+\frac{e^2}{c^2}\frac{r^2\mathcal{B}^2(r)}{4}\bigg]+V^*_{\rm KS}
    	\end{split}
    \end{equation}
\end{widetext}
where $\mathcal{B}(r)=\frac{1}{\pi r^2}\int_0^r 2\pi r' B^*(r')dr'$. Applying Eq.~(\ref{variablesepperation}) yields:
\begin{widetext}
\begin{equation}
	\begin{split}
		\frac{\partial }{\partial r}\psi_\alpha &=\frac{\partial }{\partial r} {R}_{\alpha} \frac{\exp(-im_{\alpha}\theta)}{\sqrt{2\pi r}}-{R}_{\alpha}\exp(-im_{\alpha}\theta) \frac{1}{2\sqrt{2\pi r^3}} \\
		\frac{\partial^2 }{\partial r^2}\psi_\alpha &=\frac{\partial^2 {R}_{\alpha} }{\partial r^2} \frac{\exp(-im_{\alpha}\theta)}{\sqrt{2\pi r}}-\frac{1}{\sqrt{2\pi r^3}}\frac{\partial {R}_{\alpha}}{\partial r} \exp(-im_{\alpha}\theta) +\frac{3}{4\sqrt{2\pi r^5}}{R}_{\alpha}\exp(-im_{\alpha}\theta)
	\end{split}
\end{equation}    
\end{widetext}
using $\mathcal H^* \psi_\alpha = \epsilon_\alpha \psi_\alpha $ and setting $l^2=\frac{\hbar c}{eB}=1$, the KS equation becomes
\begin{widetext}
\begin{equation}
	\frac{1}{2m^*}[-\hbar^2(\frac{\partial^2 }{\partial r^2}-\frac{1}{r}\frac{\partial }{\partial r}+\frac{3}{4r^2} +\frac{1}{r}\frac{\partial }{\partial r}-\frac{1}{2r^2}-\frac{m_\alpha^2}{r^2})+\frac{e^2}{c^2}\frac{r^2 \mathcal B^2(r)}{4}-\frac{\hbar e }{c}m_\alpha \mathcal B(r)]{R}_{\alpha}(r)+V^*_{\rm KS}{R}_{\alpha}(r)=\epsilon_\alpha {R}_{\alpha}(r) 
\end{equation}
\end{widetext}
using $\frac{1}{2}\hbar \omega_B l^2=\frac{\hbar^2}{2m^*}$, we obtain:
\begin{widetext}
\begin{equation}
	\begin{split}
		&\frac{1}{2}\hbar \omega_B [-\frac{\partial^2 }{\partial (r/l)^2}+\frac{m_\alpha^2-1/4}{(r/l)^2}+\frac{(r/l)^2 \mathcal B^2(r)}{4B^2}-m_\alpha \frac{\mathcal B(r)}{B}]{R}_{\alpha}(r)+V^*_{\rm KS}{R}_{\alpha}(r)=\epsilon_\alpha {R}_{\alpha}(r) \\
	\end{split}
\end{equation}
\end{widetext}
where $\hat{L}_z=-i\hbar \frac{\partial }{\partial \theta}$ and $\langle L_z \rangle = -m_\alpha \hbar$. It is important to note that $\langle L_z \rangle = \pm  m_\alpha \hbar$ depends on the definition of the wave function $\psi_{\alpha}(\vec{r})=\frac{{R}_{\alpha}(r)}{\sqrt{2\pi r}}\exp(\pm im_{\alpha}\theta)$. The radial wave function ${R}_{\alpha}$ satisfies the one-dimensional equation:
\begin{equation}\label{ksStar1dreduced}
  \tilde{\mathcal H}^*_{m_\alpha}(\bar{r}){R}_{\alpha}(\bar{r})=\epsilon_{\alpha}{R}_{\alpha}(\bar{r})
\end{equation}
where $\bar{r}=r/l$, and the Hamiltonian is:
 
\begin{eqnarray}
\label{1DksStarpolarform} 
		&&\tilde{\mathcal H}^*_{m_\alpha}(\bar{r})=V^*_{\rm KS}+T^*   \\
        &=& V^*_{\rm KS}+\frac{1}{2} \hbar \omega_B [-\frac{\partial^2}{\partial\bar{r}^2}+\frac{m_\alpha^2-\frac{1}{4}}{\bar{r}^2}-m_\alpha\frac{\mathcal{B}(\bar{r})}{B}+\frac{\mathcal{B}^2(\bar{r})}{B^2}\frac{\bar{r}^2}{4}] \nonumber
\end{eqnarray}
 The potential $V_T^*$ is given by:
\begin{eqnarray}\label{DeltaTStarDef}
	&&V^*_{\rm T}(\vec{r})=\sum_{\alpha}c_{\alpha}\langle\psi_{\alpha}|\frac{\delta T^*}{\delta \rho(\vec{r})}|\psi_{\alpha}\rangle \nonumber \\
    &=&\frac{1}{2}\hbar w_B \sum_{\alpha}c_{\alpha}\int d r' R_\alpha(r') \left( \frac{4m_\alpha  l_B^2  }{r'^2}- 2\frac{\mathcal{B}(r')}{ B } \right) \nonumber \\
    &&\theta(r'-r) R_\alpha(r')  
\end{eqnarray} 
 the vector potential and its functional derivative are:
 \begin{eqnarray} 
		A^*(r) &=&\frac{B}{2\pi r}\int (1-4\pi \rho l_B^2)d^2r' \nonumber\\
		\frac{\partial A^*(r)}{\partial \rho(r'')}&=&-\frac{\phi_0}{\pi r }\theta(r-r'')  
\end{eqnarray}
 where $\theta(r)$ is the step function. Eq.~(\ref{1DksStarpolarform}) can be solved numerically for each angular momentum $m_\alpha\neq 0$ using the finite-difference method. 

\section{Details of $m_\alpha=0$ Subspace}
\label{m0space}
For $m_{\alpha}=0$, we alternatively use a basis expansion method. The matrix form of $\tilde{\mathcal H}_{m_0}^*$ can be obtained using the basis set $\textbf{H}_0=\{\eta_{n,m=0}({\vec{r}},{\mathcal{B}_0}), n=0,1,\ldots,N_{\rm L}\}$ in the angular momentum $m=0$ subspace, where the basis $\eta_{n,m}({\vec{r}},{\mathcal{B}_0})$ are LL wave functions defined in Eq.~(\ref{LandauWave}) with $l_{\mathcal{B}_0}$, and $N_{\rm L}$ is the cutoff for $\Lambda$L, which is fixed at $N_{\rm L}=30$. 

We now describe how $l_{\mathcal{B}_0}$ is determined. In the $\textbf{H}_0$ subspace, we can express the matrix elements $\mathcal{H}^*_{(n',n)}$ as:
\begin{eqnarray} 
		\mathcal{H}^*_{(n',n)}&=&T^{*}_{(n',n)}+V^{*}_{\text{KS}{(n',n)}}\nonumber\\
 	T^{*}_{(n',n)}&=&\langle\eta_{n',0}({\vec{r}},{\mathcal{B}_0})|T_0+V_0|\eta_{n,0}({\vec{r}},{\mathcal{B}_0})\rangle \nonumber\\
 	&=&\hbar \omega_{\mathcal{B}_0}(n+\frac{1}{2})\delta_{n'n}+\langle\eta_{n',0}({\vec{r}},{\mathcal{B}_0})|V_0|\eta_{n,0}({\vec{r}},{\mathcal{B}_0})\rangle \nonumber\\
	V^{*}_{{\textrm{KS}}{(n',n)}}&=&\langle\eta_{n',0}({\vec{r}},{\mathcal{B}_0})|V^{*}_{\rm{KS}}|\eta_{n,0}({\vec{r}},{\mathcal{B}_0})\rangle
\end{eqnarray}
where
\begin{equation}
	\begin{split}
		T_0 &= \frac{1}{2}\hbar \omega_{\mathcal{B}_0}[\nabla_{{r/l_{\mathcal{B}_0}}}^2+\frac{\hat{L}_{z}}{\hbar}+\frac{(r/l_{\mathcal{B}_0})^2}{4}] \\
        V_0 &= \frac{1}{2}\hbar \omega_{\mathcal{B}_0}[\frac{\mathcal{B}({r})-\mathcal{B}_{0}}{\mathcal{B}_{0}}\frac{\hat{L}_{z}}{\hbar}+ \frac{1}{4}\frac{\mathcal{B}^2({r})-\mathcal{B}^2_{0}}{\mathcal{B}^2_{0}}(r/l_{\mathcal{B}_0})^2]
	\end{split}
\end{equation}
where $\omega_{\mathcal{B}_0}=\frac{e\mathcal{B}_{0}}{m^*c}$. After diagonalizing $\mathcal{H}^*_{(n',n)}$, we obtain the eigenfunctions $c_{n',n}$ and eigenvalues $\epsilon_{n,0}$. The single CF orbital with $m_\alpha=0$ can then be expressed as a linear superposition of these eigenfunctions:
\begin{equation}
   \psi_{n,0}(\vec{r})=\sum_{n'=0}^{N_{\rm L}} c_{n',n}\eta_{n',0}(\vec{r}, {\mathcal{B}_0})
\end{equation}
a reasonable choice for $\mathcal{B}_0$ is given by:
\begin{equation}\label{m0approxiamtion}
\hbar\frac{e\mathcal{B}_{0}}{m^*c} =\frac{\Delta_{m=-1}+\Delta_{m=1}}{2}
\end{equation}
where $\Delta_{m=\pm1}$ represents the average cyclotron energy gap between the lowest two energy levels in the $m=\pm 1$ angular momentum subspace, as obtained using the finite difference method. This approximation is valid because orbitals with $m=0$ are spatially overlapping with those of $m=\pm1$ and therefore experience a similar effective magnetic field. 
\section{Iterative Procedure of DFT}
\label{iterative}
The iterative procedures for KS-DFT is as follows:
\begin{itemize}
  \item (i) \textbf{Initialization:} Start with an initial density $\rho_{\text{in}}$, typically set to the  background density.
  \item (ii) \textbf{Calculation:} Determine $T^*$ and $V^*_{\rm KS}(\vec{r})$ for corresponding orbitals, and diagonalize the Hamiltonian to obtain the KS orbitals. These orbitals yield an output density $\rho_{\rm out}=\sum_{n,m} c_{n,m}|\psi_{n,m}|^2$, where $c_{n,m}$ is the occupation number for the KS orbitals labeled by $\{n,m\}$.
  \item (iii)\textbf{Convergence Check:} Calculate the relative difference $\Delta =  \frac{\int d\vec{r} |\rho_{\rm out}-\rho_{\rm in}|}{N}$. The density $\rho_{\rm out}$ is considered converged if $\Delta<10^{-5}$.
  \item (iv) \textbf{Update:} If convergence is not achieved, update the input density $\rho'_{\rm in}$ by mixing the output density with the previous input density: $\rho'_{\rm in} = \lambda \rho_{\rm in}+(1-\lambda)\rho_{\rm out}$, where the mixing coefficient $\lambda= 0.95$ is used. Repeat the iterative process until convergence is achieved.
\end{itemize}

\bibliography{biblio_fqhe.bib}

\end{document}